\documentclass[referee]{raa}

\usepackage{graphicx,times}
\usepackage{natbib}
\usepackage{url}
\usepackage{amssymb,amsmath}
\usepackage{tikz,xcolor}
\usepackage{setspace}

\usepackage[pagebackref=true]{hyperref}
\usepackage{orcidlink}

\usepackage{CJKutf8}
\newcommand{\namecn}[1]{\begin{CJK*}{UTF8}{gbsn}({#1})\end{CJK*}}

\bibpunct{(}{)}{;}{a}{}{,}
\begin{document}

\title{The 3D Architecture of a pair of 6:1 Resonant Brown Dwarfs around the Naked-eye star $\nu$ Ophiuchi}

\volnopage{ {\bf 20XX} Vol.\ {\bf X} No. {\bf XX}, 000--000}
\setcounter{page}{1}

\author{
Tianshenhong Sang \namecn{桑田甚红}\orcidlink{0009-0006-8208-2311} \inst{1,2} 
\and  Guang-Yao Xiao \namecn{肖光耀}\orcidlink{0000-0001-6753-4611} \inst{3} 
\and  Ying-Yi Cao \namecn{曹映怡} \inst{4}
\and  Huan-Yu Teng  \namecn{滕环宇}\orcidlink{0000-0003-3860-6297} \inst{1,*}  
\and  Yu-Juan Liu  \namecn{刘玉娟}\orcidlink{0009-0008-3430-1027} \inst{1,*}\footnotetext{$^{*}$Corresponding Authors} 
\and  Wei Wang  \namecn{王炜}\orcidlink{0000-0002-9702-4441} \inst{1} 
\and  Fan Liu  \namecn{刘凡}\orcidlink{0000-0003-4794-6074} \inst{1} 
\and  Fei Zhao  \namecn{赵斐}\orcidlink{0000-0003-4276-1767} \inst{1} 
\and  Meng Zhai  \namecn{翟萌}\orcidlink{00000-0003-1207-3787} \inst{1} 
\and  Fabo Feng  \namecn{冯发波}\orcidlink{0000-0001-6039-0555} \inst{3}  
\and  Shang-Fei Liu \namecn{刘尚飞} \inst{5}
}

\institute{
National Astronomical Observatories, Chinese Academy of Sciences, Beijing 100101; {\it lyj@nao.cas.cn} {\it hyteng@bao.ac.cn}\\
\and School of Astronomy and Space Science, University of Chinese Academy of Sciences, Beijing 100049, China\\
\and State Key Laboratory of Dark Matter Physics, Tsung-Dao Lee Institute \& School of Physics and Astronomy, Shanghai Jiao Tong University, Shanghai 201210, China\\
\and Department of Earth and Planetary Sciences, The University of Hong Kong, Pokfulam Road, Hong Kong, China  \\
\and School of Physics and Astronomy, Sun Yat-sen University Zhuhai Campus, 2 Daxue Road,
Tangjia, Zhuhai 519082, Guangdong Province, China \\
\vs \no
{\small Received 20XX Month Day; accepted 20XX Month Day}
}

\abstract{
We present a revisiting study of the brown dwarf pair orbiting the naked-eye ($V=3.3$) K-giant $\nu$~Ophiuchi, located only 44\,pc from our Solar system. By jointly analysing archival radial-velocity measurements together with astrometric data from \textit{Hipparcos} and the \textit{Gaia} second and third data releases, we determine the three-dimensional architecture of the system and robustly constrain the masses of both companions. We find brown dwarf masses of $m_{\mathrm{b}} = 24.2^{+6.4}_{-2.8}\,M_{\mathrm{J}}$ and $m_{\mathrm{c}} = 26.8^{+4.3}_{-2.9}\,M_{\mathrm{J}}$. The mathematical constraint, derived from the posterior distribution of the mutual inclination based on MCMC samples, yields a mutual inclination of $\psi_{\mathrm{bc}}=46^{+27}_{-24}\!\,^{\circ}$, while direct calculations based on the maximum a posteriori and posterior median orbital parameters yield values of $\sim$$10^{\circ}$ and $\sim$$20^{\circ}$, respectively. Resonance analysis indicates that the two companions can still be trapped in a 6:1 mean-motion resonance in the maximum a posteriori configuration. To place an upper limit for the mutual inclination, dynamical stability analysis over a 1~Myr timescale further constrains it to be no larger than $\sim$$15^{\circ}$. Systems hosting brown dwarf pairs are rare, yet they provide important constraints on theories of planetary formation and dynamical evolution. Current detections suggest that brown dwarf pairs preferentially reside at large separations from their host stars and are more common in less mature systems. This supports a star-like formation pathway via gravitational instability in disk.
}

\authorrunning{T. Sang et al. }            
\titlerunning{The 3D Architecture of $\nu$ Ophiuchi system}  
\maketitle

%
\section{Introduction}\label{sect:intro}  
The radial-velocity (RV; or Doppler) method has been employed to search for planetary systems for more than 30~years \citep{Mayor1995}. While it efficiently derives planetary orbits even with sparse temporal sampling, it suffers from an inherent degeneracy between the orbital inclination and the companion mass. This degeneracy can be readily broken for transiting planets, which pass through the observer’s line of sight, but this imposes a requirement on the orbital orientation. In addition, long-period companions have a smaller transit probability, and RV measurements alone are far less sensitive for them. 
An alternative and efficient approach for such systems is astrometry (e.g., \citealt{Fischer2014} ). The first example of breaking this degeneracy is the GJ~876 system, presented by \citet{Benedict2002} through a joint analysis of astrometric measurements from the \textit{Hubble} Space Telescope Fine Guidance Sensor and archival RV data.

More recently, methodologies based on jointly constraining proper-motion data from different astrometric facilities such as \textit{Hipparcos} \citep{Leeuwen2007}, \textit{Gaia} \citep{Gaia2018,Gaia2023} and simulated observations based on the \textit{Closeby Habitable Exoplanet Survey} (\textit{CHES}; \citealt{Ji2022,Bao2024,Bao2024AJ....167..286,Ji2024,Tan2024}) with RV data have been independently developed by several groups (e.g., \citealt{Feng2019,Xuan2020,Venner2021,Brandt2021b,Kervella2022,Feng2025,Huang2025}). As a result, an increasing number of planetary systems, including multiple-planet systems, have been characterised with their orbital inclinations and true masses determined (e.g., \citep{Li2021,GMBrandt2021,Feng2022,Xiao2023,An2025}). Within current observational capabilities, precise astrometric measurements can be obtained for planetary and substellar companion systems within approximately 100~pc of the Solar neighbourhood. Three-dimensional characterisation of planetary systems facilitates population-level studies of cold giant companions and provides important insights into their formation scenarios (e.g., \citealt{Feng2022,Xiao2023,An2025,Ji2022,Ji2024}).

The star $\nu$~Oph (HD~163917) is a naked-eye ($V=3.3$) K-type giant star located at a distance of 44~pc and hosts two brown dwarf companions. The host star has a mass of $2.6\,M_{\odot}$ and an estimated age of $\sim$650~Myr \citep{Teng2023}. The two brown dwarfs were previously reported to have minimum masses of approximately $22\,M_{\mathrm{J}}$ and $25\,M_{\mathrm{J}}$, and are likely locked in a 6:1 mean-motion resonance with orbital periods of $\sim$530 and $\sim$3180~days \citep{Quirrenbach2019,Teng2023,Spaeth2025}. Previous analyses of resonances in giant-companion pairs have largely assumed coplanar or nearly coplanar configurations, while direct observational constraints on their true mutual inclinations remain blank. Given the proximity of $\nu$~Oph and the substantial masses of its companions in the substellar regime, constraining the orbital inclinations of both companions becomes feasible using astrometric data from \textit{Hipparcos} and the \textit{Gaia} second and third data releases (DR2 and DR3). This joint analysis enables the geometric distinction between prograde and retrograde configurations of the two companions \citep{Feng2025,Xiao2025}. Consequently, $\nu$~Oph provides an excellent laboratory for directly probing the three-dimensional architecture of resonant giant-companion pairs.

This paper is organised as follows. In Section~\ref{sect:data}, we describe the data sets and our methodology for the joint analysis of RV and astrometric measurements. In Section~\ref{sec:results}, we present the resulting constraints, and in Section~\ref{sect:discussion}, we discuss the implications of our findings and summarise this work.

\section{Data, Methods, and Analysis}\label{sect:data}
The precise RV measurements of $\nu$~Oph span more than 22~years, from 2000 to 2022 \citep{Sato2012,Quirrenbach2019,Teng2023,Spaeth2025}. In total, more than 228 RV data points were obtained with the Hamilton Spectrograph \citep{Vogt1987} mounted on the 0.6\,m Coud\'e Auxiliary Telescope at UCO/Lick Observatory, the High Dispersion Echelle Spectrograph (HIDES; \citealt{Izumiura1999}) mounted on the 1.88\,m telescope at Okayama Astrophysical Observatory, and the CARMENES spectrograph installed on the 3.5\,m telescope at Calar Alto Observatory. We retrieved the original RV data from \citet{Quirrenbach2019}, \citet{Teng2023}, and \citet{Spaeth2025}. For the CARMENES observations, only the visible-channel RVs were included in this analysis, as the simultaneous near-infrared data have a lower signal-to-noise ratio. In addition, although ten near-infrared RV measurements were obtained with the CRIRES spectrograph on the Very Large Telescope at Paranal, these data were not included in our analysis owing to their limited orbital phase coverage.

The astrometric data were obtained from both \textit{Hipparcos} \citep{Leeuwen2007} and \textit{Gaia} DR2 and DR3 \citep{Gaia2018,Gaia2023}. Because the $\nu$~Oph system hosts two companions, relying on absolute astrometry from only \textit{Hipparcos} and \textit{Gaia} DR3 (referred to as proper-motion anomalies, PMA) would introduce degeneracies in the inferred orbital inclinations. To mitigate this issue, we simulated the \textit{Gaia} epoch astrometry using the Gaia Observation Forecast Tool\footnote{\url{https://gaia.esac.esa.int/gost/index.jsp}} (\texttt{GOST}), and incorporated additional information from \textit{Gaia} DR2, which is crucial for constraining the true orbital orientations of the companions. The detailed methodology is described in \citet{Feng2025}.

The construction of our joint model combining RV and astrometric data has been described in detail in previous studies (e.g., \citealt{Xiao2024,Xiao2025,Feng2025}\footnote{\url{https://github.com/gyxiaotdli/mini_Agatha}}). We note that the Keplerian model was adopted for this construction.
Given the observational uncertainties of a few $\mathrm{m}\,\mathrm{s}^{-1}$, the dominant solar-like oscillation jitter ($\sim$20 $\mathrm{m}\,\mathrm{s}^{-1}$; \citealt{Bedding2011,Teng2023}), and other sources of stellar jitter at the level of a few $\mathrm{m}\,\mathrm{s}^{-1}$ \citep{Teng2023}, neglecting the interactions between the two companions in the orbital fit is acceptable within the current observational baseline. Under these conditions, the Keplerian model provides a practical and efficient description of the RV data. Here we briefly summarise the key elements. The orbital parameters directly fitted in our framework include the logarithm of the orbital period $\ln P$, the logarithm of the RV semi-amplitude $\ln K$, the eccentricity vector components $\sqrt{e}\cos\omega$ and $\sqrt{e}\sin\omega$ (where $\omega$ denotes the argument of periastron of the stellar reflex motion), the mean anomaly at the epoch of the first RV observation $M_{0}$, the orbital inclination $i$, the longitude of the ascending node $\Omega$, and five astrometric offsets of the barycentre relative to \textit{Gaia} DR3, namely $(\Delta\alpha^{*},\,\Delta\delta,\,\Delta\mu_{\alpha^{*}},\,\Delta\mu_{\delta},\,\Delta\varpi)$. The fitted parameters and their adopted priors are listed in Table~\ref{tab:orbit}. Derived quantities, including the semi-major axis $a$ and the companion mass $m_{\mathrm{p}}$, are computed from these fitted parameters. The posterior distributions are sampled using a parallel-tempering Markov Chain Monte Carlo (MCMC) sampler, \texttt{ptemcee} \citep{Vousden2016}, with 30 temperatures, 100 walkers, and 80{,}000 steps per chain, of which the first 20{,}000 steps are discarded as burn-in. We quantified the convergence of the walkers using the Gelman--Rubin diagnostic ($\hat{R}$; \citealt{Gelman2013}). After 80{,}000 steps, the maximum $\hat{R}$ value was below the commonly adopted convergence threshold of 1.01, indicating that the chains had converged well.

\section{Results}\label{sec:results}
\begin{figure}[t]
\centering
\includegraphics[width=\textwidth]{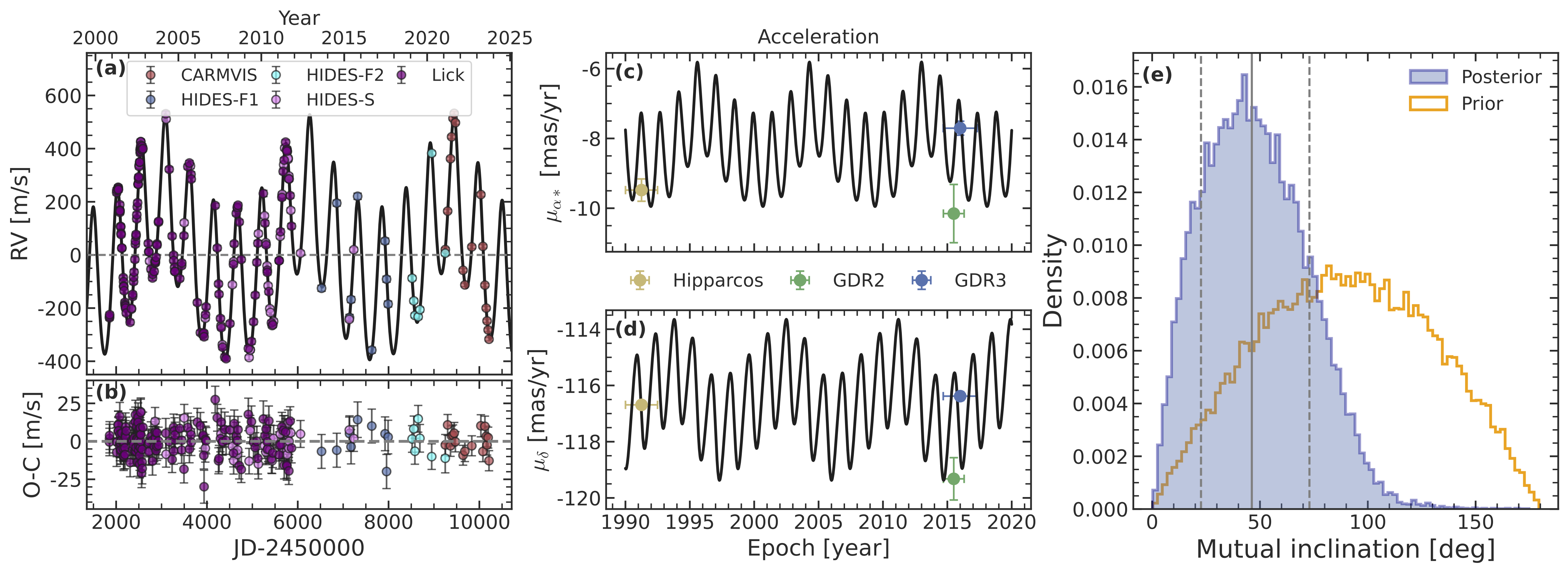}
\caption{Best-fit orbital solution for the $\nu$~Oph system. Panels~(a) and~(b) show the best-fit radial-velocity model, with measurements from different instruments indicated by different colours. A zoomed-in plot of panels (a) and (b) are given in Fig. \ref{fig:rv} in Appendix.} Panels~(c) and~(d) present the best-fit solution to the astrometric acceleration. Panel~(e) displays the posterior distribution of the mutual inclination between companions~b and~c derived from the orbital fit.
\label{fig:rvast}
\end{figure}

The joint analysis of precise RV and astrometric data shows that key orbital parameters, including the orbital periods, eccentricities, and semi-major axes, are in good agreement with recent studies \citep{Quirrenbach2019,Teng2023,Spaeth2025}, mostly within 1$\sigma$ level.  In addition, the orbital inclinations of both the inner and outer brown dwarf companions are constrained. 

Constraints on the remaining orbital parameters for both companions are listed in Table~\ref{tab:orbit}, and the values in the brackets are the values at maximum posterior. In particular, for the inner brown dwarf, we derive an orbital period of $P_{\mathrm{b}} = 529.91 \pm 0.06\,(529.99)$~days, a semi-major axis of $a_{\mathrm{b}} = 1.77^{+0.07}_{-0.06}\,(1.78)$~au, and an eccentricity of $e_{\mathrm{b}} = 0.124 \pm 0.003\,(0.124)$. The inferred orbital inclination is $i_{\mathrm{b}} = 84^{+36}_{-32}\,(65)$~degrees, corresponding to a mass of $m_{\mathrm{b}} = 24.2^{+6.4}_{-2.8}\,(22.8)\,M_{\mathrm{J}}$. 
For the outer brown dwarf, we obtain an orbital period of $P_{\mathrm{c}} = 3175.3 \pm 4.3\,(3175.2)$~days, a semi-major axis of $a_{\mathrm{c}} = 5.85^{+0.21}_{-0.23}\,(5.86)$~au, and an eccentricity of $e_{\mathrm{c}} = 0.185 \pm 0.006\,(0.186)$. The corresponding orbital inclination is $i_{\mathrm{c}} = 64^{+16}_{-12}\,(60)$~degrees, yielding a mass of $m_{\mathrm{c}} = 26.8^{+4.3}_{-2.9}\,(25.8)\,M_{\mathrm{J}}$. 

Our best-fit model yields $\chi_{\rm red,RV}^2 = 1.005$, indicating a goodness of fit and supporting the validity of the adopted RV model. To further justify the use of the Keplerian approximation, we calculated the difference between the Keplerian and non-Keplerian models. We integrated the orbits using the maximum-a-posteriori orbital parameters with the \texttt{rebound} package \citep{Rein2012} and the \texttt{IAS15} integrator \citep{Rein2015a}, which provides high-accuracy orbital evolution. As shown in Fig.~\ref{fig:rv_diff}, we find a difference of about 10~m~s$^{-1}$ after 6000 days and a maximum difference of $\sim$15~m~s$^{-1}$ within the observational baseline of $\sim$8500 days. Considering the expected stellar jitter of more than $\sim$20~m~s$^{-1}$, additional noise sources at the level of a few m~s$^{-1}$, and observational uncertainties of a few m~s$^{-1}$, the difference between the Keplerian and non-Keplerian models is likely to be masked by the jitter and observational noise. Therefore, a Keplerian model with an additional jitter term remains a valid description of the current RV data. We note, however, that if the observational baseline becomes sufficiently long for the accumulated non-Keplerian effect to become significant compared with the jitter level, the Keplerian approximation would no longer be adequate. Furthermore, in order to examine the additional signals after subtracting the orbital signals, we performed a Generalised Lomb--Scargle periodogram analysis \citep{Zechmeister2009} and confirmed the absence of any additional long-term periodicity in the RV residuals as shown in Fig.~\ref{fig:gls_residuals}. Additional moderate signals at periods of a few tens of days may be attributed to unresolved or unknown stellar variability and/or instrumental noise \citep{Teng2023, Spaeth2025}.

Our results confirm that both companions reside firmly in the brown dwarf mass regime ($13\,M_{\mathrm{J}} < m_{\mathrm{p}} < 80\,M_{\mathrm{J}}$). The inclination constraint for the inner brown dwarf exhibits a relatively large uncertainty of $\sim$35~degrees, which arises from the small angular separation between the inner companion and the host star. With the orbital inclinations and longitudes of the ascending nodes determined for both companions, the mutual inclination $\psi_{\mathrm{bc}}$ can be derived by
\begin{equation}
    \cos \psi_{\mathrm{bc}} = \cos i_{\mathrm{b}} \cos i_{\mathrm{c}} + \sin i_{\mathrm{b}} \sin i_{\mathrm{c}} \cos \Delta \Omega_{\mathrm{bc}},
    \label{eq:mutual}
\end{equation}
where $\Delta \Omega_{\mathrm{bc}} = \Omega_{\mathrm{b}} - \Omega_{\mathrm{c}}$. From the posterior distributions, we obtain a mutual inclination of $\psi_{\mathrm{bc}} = 46^{+27}_{-24}\,(42)$~degrees, indicating a likely inclined configuration between the two companions at approximately the $2\sigma$ level. We note that the value of 42~degrees corresponds to the maximum of the posterior distribution for $\psi_{\mathrm{bc}}$, and does not equal the $\sim$10~degrees or $\sim$21~degrees value obtained by combining the individual maximum a posteriori or posterior median estimates of $i_{\mathrm{b}}$, $i_{\mathrm{c}}$, $\Omega_{\mathrm{b}}$, and $\Omega_{\mathrm{c}}$, respectively. This is due to the asymmetric posterior distributions, especially for  $i_{\mathrm{b}}$.

At present, our methodology and results serve as a temporary solution until the \textit{Gaia} fourth data release becomes available (expected no earlier than 2026). In particular, the forthcoming release is expected to provide substantially tighter constraints on the inner brown dwarf ($P \simeq 530\,\mathrm{d}$), as \textit{Gaia} achieves its highest astrometric sensitivity for orbital periods between roughly hundreds of days (e.g., \citealt{Lammers2026}).

\begin{table}[]
\footnotesize
\centering
\begin{tabular}{lcccc}
\hline \hline
Parameter & Unit & Meaning & RV+HG23 & Prior${\rm ^c}$ \\
\hline
$\ln P_{\rm{b}}$&day&Orbital period & ${6.27270}\pm{0.00011}\,(6.27163)$& $\mathcal{U}(-1,16)$\\
$\ln K_{\rm{b}}$&m s$^{-1}$&RV semi-amplitude&${5.6654}\pm{0.031}\,(5.6647)$& $\mathcal{U}(-6,6)$\\
$\sqrt{e_{\rm{b}}}\cos\omega_{\rm{b}}$&---&Combination of Eccentricity&${0.3464}_{-0.0048}^{+0.0046}\,(0.34714)$& $\mathcal{U}(-1,1)$\\
$\sqrt{e_{\rm{b}}}\sin\omega_{\rm{b}}$&---&and Argument of periapsis${\rm ^a}$&${0.0615}\pm{0.0080}\,(0.06122)$& $\mathcal{U}(-1,1)$\\
$M_{0,\rm{b}}$&deg&Mean anomaly at 1st epoch &${235.4}\pm{1.3\,(235.4)}$& $\mathcal{U}(0,360)$\\
$i_{\rm{b}}$&deg&Inclinations&${84}_{-32}^{+36}\,(65)$ & $\cos i$-$\mathcal{U}(-1,1)$\\
$\Omega_{\rm{b}}$&deg&Longitude of ascending node&${316}_{-60}^{+27}\,(325)$ & $\mathcal{U}(0,360)$\\
\hline
$\ln P_{\rm{c}}$&day&Orbital period & ${8.0631}_{-0.0014}^{+0.0013}\,(8.0627)$& $\mathcal{U}(-1,16)$\\
$\ln K_{\rm{c}}$&m s$^{-1}$&RV semi-amplitude&${5.1728}_{-0.0068}^{+0.0065}\,(5.1727)$& $\mathcal{U}(-6,6)$\\
$\sqrt{e_{\rm{c}}}\cos\omega_{\rm{c}}$&---&Combination of Eccentricity&${0.4265}_{-0.0070}^{+0.0067}\,(0.4270)$& $\mathcal{U}(-1,1)$\\
$\sqrt{e_{\rm{c}}}\sin\omega_{\rm{c}}$&---&and Argument of periapsis${\rm ^a}$&${0.058}_{-0.013}^{+0.012}\,(0.061)$& $\mathcal{U}(-1,1)$\\
$M_{0,\rm{c}}$&deg&Mean anomaly  at 1st epoch   &${223.1}_{-1.6}^{+1.7}\,(223.1)$& $\mathcal{U}(0,360)$\\
$i_{\rm{c}}$&deg&Inclinations&${64}_{-12}^{+16}\,(60)$& $\cos i$-$\mathcal{U}(-1,1)$\\
$\Omega_{\rm{c}}$&deg&Longitude of ascending node&${323}_{-46}^{+21}\,(334)$  & $\mathcal{U}(0,360)$\\
\hline
$\Delta \alpha*$&mas&$\alpha*$ offset&${0.27}_{-0.57}^{+0.40}\,(0.37)$& $\mathcal{U}(10^{-6},10^{6})$\\
$\Delta \delta$&mas&$\delta$ offset&${-1.45}_{-0.28}^{+0.37}\,(-1.51)$& $\mathcal{U}(10^{-6},10^{6})$\\
$\Delta \mu_{\alpha*}$&mas\,yr$^{-1}$&$\mu_{\alpha*}$ offset &${0.434}_{-0.015}^{+0.016}\,(0.431)$& $\mathcal{U}(10^{-6},10^{6})$\\
$\Delta \mu_\delta$&mas\,yr$^{-1}$&$\mu_\delta$ offset&${0.147}\pm{0.009}\,(0.149)$& $\mathcal{U}(10^{-6},10^{6})$\\
$\Delta \varpi$&mas&$\varpi$ offset&${0.56}\pm{0.18}\,(0.58)$& $\mathcal{U}(10^{-6},10^{6})$\\
\hline
{$M_{\star}$}& {$M_{\odot}$} & {Stellar mass${\rm ^b}$} &{---}& {$\mathcal{N}(2.6,0.3)$}\\
\hline
$J^{\rm CARMVIS}$&m\,s$^{-1}$&RV jitter for CARMVIS&${7.4}_{-1.5}^{+2.0}\,(7.0)$&$\mathcal{U}(0,10^{6})$\\
$J^{\rm HIDES-S}$&m\,s$^{-1}$&RV jitter for HIDES-S&${7.1}_{-1.0}^{+1.1}\,(7.0)$&$\mathcal{U}(0,10^{6})$\\
$J^{\rm HIDES-F1}$&m\,s$^{-1}$&RV jitter for HIDES-F1&${10.4}_{-2.7}^{+4.1}\,(9.5)$&$\mathcal{U}(0,10^{6})$\\
$J^{\rm HIDES-F2}$&m\,s$^{-1}$&RV jitter for HIDES-F2&${10.5}_{-3.5}^{+5.7}\,(8.8)$&$\mathcal{U}(0,10^{6})$\\
$J^{\rm Lick}$&m\,s$^{-1}$&RV jitter for Lick&${7.4}\pm{0.7}\,(7.4)$&$\mathcal{U}(0,10^{6})$\\
$J^{\rm Hipparcos}$&mas&Jitter for Hipparcos&${0.73}_{-0.14}^{+0.15}\,(0.74)$&$\mathcal{U}(0,10^{6})$\\
$S^{\rm Gaia}$&---&Error inflation factor&${1.25}\pm{0.08}\,(1.25)$&$\mathcal{N}(1,0.1)$\\
\hline 
$P_{\rm{b}}$&d&Orbital period &${529.91}\pm{0.06}\, (529.99)$& (Derived)\\
$a_{\rm{b}}$&au&Semi-major axis${\rm ^b}$ &${1.77}_{-0.06}^{+0.07}\, (1.78)$& (Derived) \\
$e_{\rm{b}}$&---&Eccentricity&${0.124}\pm{0.003}\, (0.124)$& (Derived)\\
$m_{\rm p,b}$&$M_{\rm J}$&Companion mass &${24.2}_{-2.8}^{+6.4}\, (22.8)$& (Derived)\\
$T_{\rm p,b}-2400000$&JD&Periapsis epoch&${51507.0}_{-2.0}^{+1.9}$& (Derived)\\
\hline
$P_{\rm{c}}$&d&Orbital period &${3175.3}\pm{4.3}\, (3175.2)$& (Derived)\\
$a_{\rm{c}}$&au&Semi-major axis${\rm ^b}$ &${5.85}_{-0.23}^{+0.21}\, (5.86)$& (Derived)\\
$e_{\rm{c}}$&---&Eccentricity&${0.185}\pm{0.006}\, (0.186)$& (Derived)\\
$m_{\rm p,c}$&$M_{\rm J}$&Companion mass &${26.8}_{-2.9}^{+4.3}\, (25.8)$& (Derived)\\
$T_{\rm p,c}-2400000$&JD&Periapsis epoch&${49885}_{-16}^{+15}$& (Derived)\\
\hline
$\psi_{\rm{bc}}$&deg&Mutual Inclination&${46}_{-24}^{+27}\,(42)$& (Derived)\\
\hline
{RMS$_{\rm RV}$} & {m s$^{-1}$} & {RMS for RV residuals} & {8.8} & {(Derived)} \\
{$\chi_{\rm RV}^2$} & {---} & {$\chi^2$ for RV } & {278.8} & {(Derived)} \\
{$\chi_{\rm red,RV}^2$} & {---} & {Reduced-$\chi^2$ for RV } & {1.005} & {(Derived)} \\
{$\chi_{\rm Hipparcos}^2$} & {---} & {$\chi^2$ for Hipparcos } & {81.82} & {(Derived)} \\
{$\chi_{\rm red,Hipparcos}^2$} & {---} & {Reduced-$\chi^2$ for Hipparcos } & {1.544} & {(Derived)} \\
{$\chi_{\rm GOST:GDR2}^2$} & {---} & {$\chi^2$ for GOST:GDR2${\rm ^d}$} & {23.56} & {(Derived)} \\
{$\chi_{\rm GOST:GDR3}^2$} & {---} & {$\chi^2$ for GOST:GDR3${\rm ^d}$} & {18.95} & {(Derived)} \\
\hline
\end{tabular}
\caption{
Parameters for $\nu$ Oph. The results are given in posterior median value with their uncertainties given by the 16\% and 84\% percentiles from the posterior distribution, and the values in the brackets are maximum a posteriori values. Here we also note: $\rm ^a$ The argument of periastron of the stellar reflex motion, differing by $\pi$ with planetary orbit, i.e., $\omega_{\rm p}=\omega+\pi$.
$\rm ^b$ The semi-major axis $a$ and planet mass $m_{\rm p}$ are derived from fitted parameters assuming the stellar mass as a Gaussian prior of $\mathcal{N}(2.6,0.3)$.
$\rm ^c$ The prior $\cos i$-$\mathcal{U}(a, b)$ is the cosine uniform distribution between $a$ and $b$, and $\mathcal{N}(a, b)$ is the Gaussian distribution with mean $a$ and standard deviation $b$.
${\rm ^d}$ The reduced-$\chi^2$ values for GOST:GDR2 and GOST:GDR3 are not applicable. We refer readers to \citet{Feng2025} for details of the methodology.
The instrumental RV offsets were marginalised in the likelihood function. The MCMC posterior distributions of key parameters are given in Fig. \ref{fig:mcmc_posterior_5} in Appendix.
}
\label{tab:orbit}
\end{table}

\section{Discussion and Summary}\label{sect:discussion}
\subsection{Resonance and stability of the $\nu$ Oph system}
The $\nu$~Oph system was previously reported to reside in a stable 6:1 mean-motion resonance \citep{Quirrenbach2019}. However, observations prior to the \textit{Gaia} era relied only on RV data, such that the orbital inclinations and true masses of both brown dwarfs (BDs) were observationally degenerate. Even dynamical fits aimed at breaking this degeneracy relied on strong assumptions, for example, coplanar or near-coplanar configurations with small mutual inclinations. 
With the advantage of high-precision astrometric measurements from \textit{Gaia}, the three-dimensional architecture of the system can now be constrained. 
In this work, our Keplerian fit revealed that the two BDs likely have a relatively large mutual inclination ($\psi =46^{+27}_{-24}\!\,^{\circ}$), which is inconsistent with the former assumption of coplanarity, motivating a re-examination of the system's resonant configuration in light of the updated three-dimensional orbital constraints.

The 6:1 resonance is associated with twenty-eight possible resonant angles, including eccentricity-type and mixed eccentricity–inclination resonant angles\citep{Murray1999}. 
The general resonant angle can be written as
\begin{equation}
    \phi = j_{1}\lambda_{1} + j_{2}\lambda_{2} + j_{3}\varpi_{1} + j_{4}\varpi_{2} + j_{5}\Omega_{1} + j_{6}\Omega_{2},
    \label{eq:resonant angle}
\end{equation}
where $\lambda$ denotes the mean longitude, $\varpi$ denotes the longitude of pericenter, $\Omega$ denotes the longitude of the ascending node, and $\Sigma_{i=1}^{6}j_{i}=0$.
For the maximum a posteriori solution, we computed all twenty-eight resonant angles and identified that six eccentricity-type resonant angles ($j_5=j_6=0$) librate about zero, while the remaining twenty-two resonant angles circulate, indicating a purely eccentricity-type resonance. For the posterior median solution, all resonant angles circulate, suggesting that the system is likely not in mean motion resonance in this configuration. Given that the maximum a posteriori solution exhibits a pure eccentricity-type resonance, with no evidence for inclination coupling, it is reasonable to adopt the mixed pericentre angle $\hat{\varpi}_{1,2}$ \citep{Sessin1984,Henrard1986,Wisdom1986,Batygin2013,Hadden2019} to construct a single representative resonant angle \citep{Dai2023} for simplicity, which can be written as
\begin{equation}
    \phi_{1,2} = p\lambda_{2} - (p-q)\lambda_{1} - q\hat{\varpi}_{1,2},
\end{equation}
where $p\!:\!q = 6\!:\!1$ is the resonance ratio, and $|p-q| = 5$ is the order of the resonance. The mixed pericenter angle $\hat{\varpi}_{1,2}$ is given by
\begin{equation}
    \hat{\varpi}_{1,2} = \arctan\!\left[
    \frac{f\,e_{1}\sin\varpi_{1} + g\,e_{2}\sin\varpi_{2}}
         {f\,e_{1}\cos\varpi_{1} + g\,e_{2}\cos\varpi_{2}}
    \right],
\end{equation}
where $f$ and $g$ are the high-order weighting coefficients, which can be computed using the \texttt{celmech} package \citep{Hadden2019}. By examining the temporal evolution of $\phi_{1,2}$, the resonant state of the system can be assessed: libration about zero indicates a resonant and dynamically stable configuration, whereas circulation over the full range $0$--$2\pi$ indicates a near-resonant, potentially chaotic state.

\begin{figure}[t]
    \centering
    \includegraphics[width=0.31\textwidth]{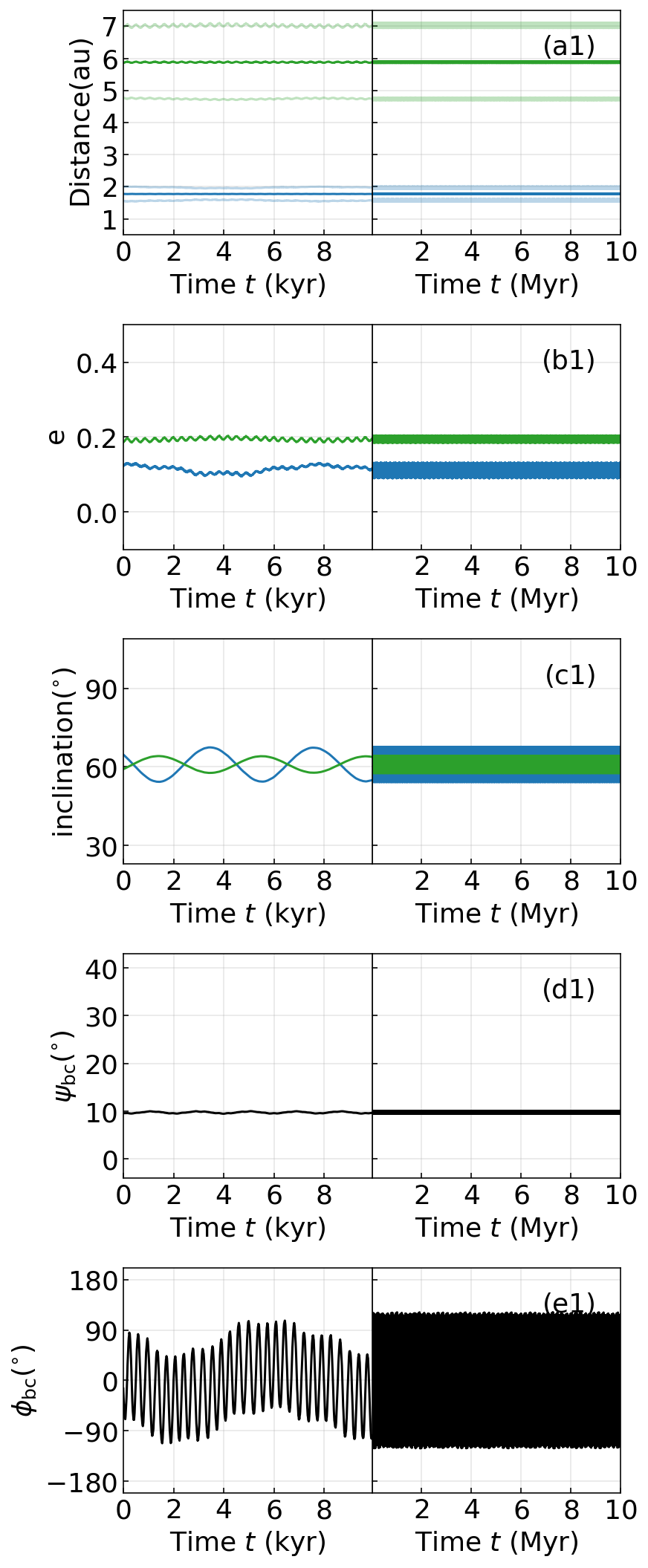}
    \includegraphics[width=0.31\textwidth]{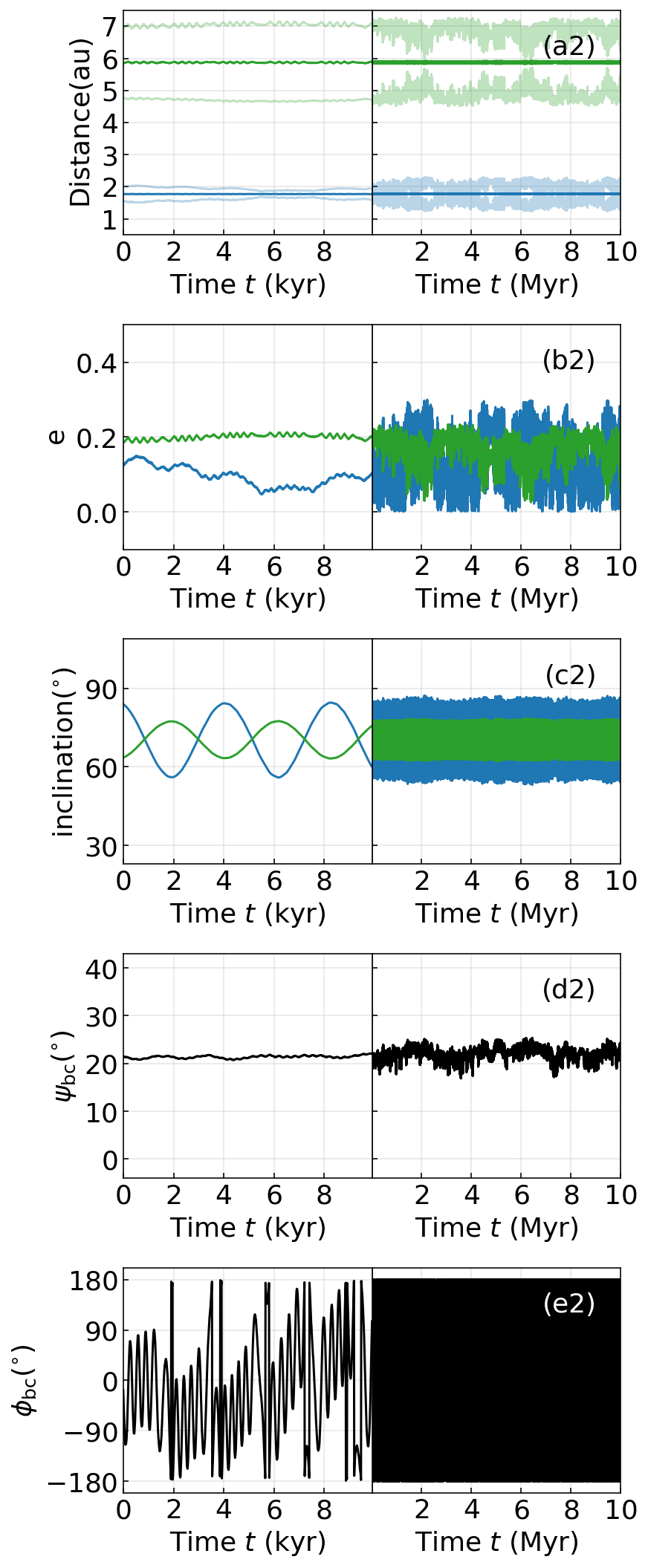}
    \includegraphics[width=0.31\textwidth]{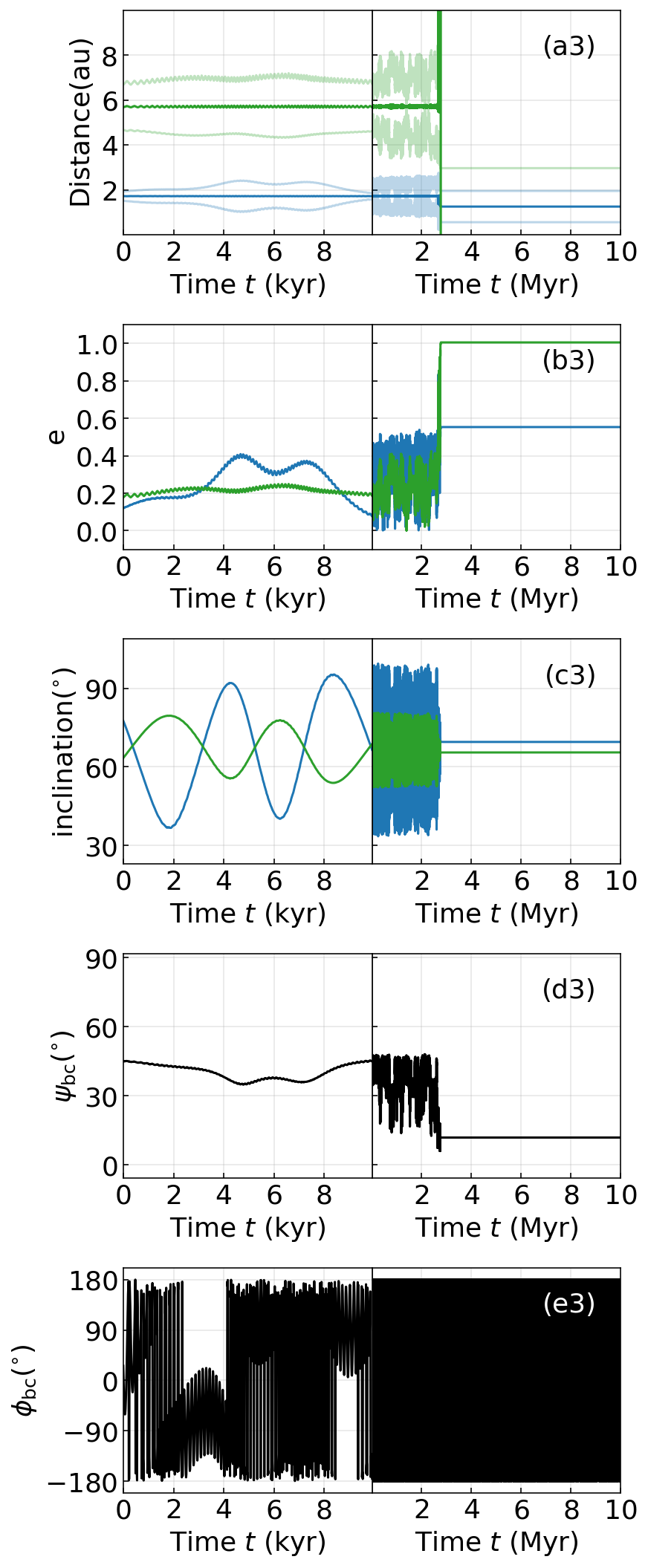}
    \caption{Ten-Myr dynamical evolution of the $\nu$~Oph system, with the left panels (a1-e1), middle panels (a2-e2), and right panels (a3-e3) showing results generated from maximum a posteriori value, posterior median value, and an MCMC sample of orbital parameters with a mutual inclination in the range $40^{\circ}$$-$$50^{\circ}$, respectively.  In each panel, a zoomed-in view of the first 10~kyr is presented on the left. From top to bottom, the panels display the orbital distances of the two companions (a1, a2, a3), the orbital eccentricities (b1, b2, b3), the orbital inclinations of the two companions (c1, c2, c3), the mutual inclination (d1, d2, d3), and the resonant angle (c1, c2, c3). In the panels (a1-3), (b1-3) and (c1-3), the blue and green curves correspond to companions~b and~c, respectively, and in the panels (a1-3), the solid curves denote the semi-major axes, while the transparent curves above and below indicate the apocentre and pericentre distances.}
    \label{fig:resonance}
\end{figure}

\begin{figure}[t]
    \centering
    \includegraphics[width=0.49\textwidth]{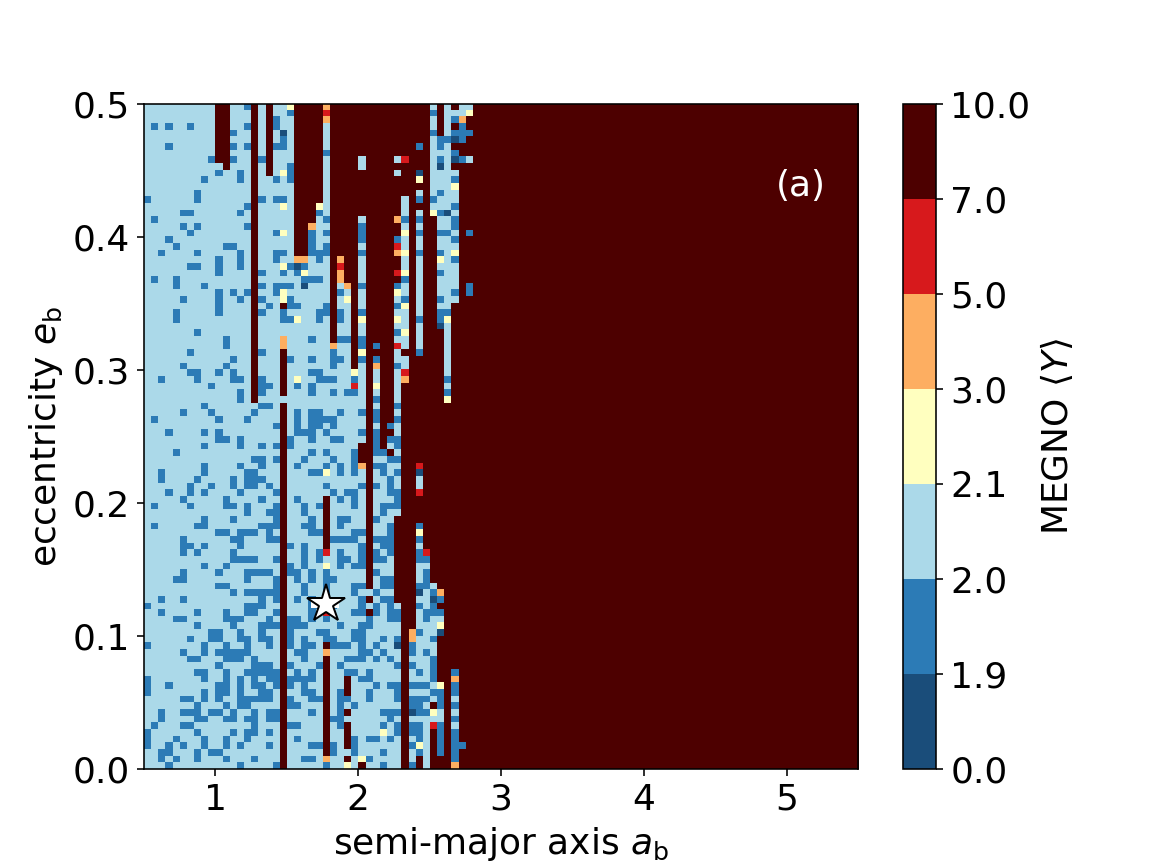}
    \includegraphics[width=0.49\textwidth]{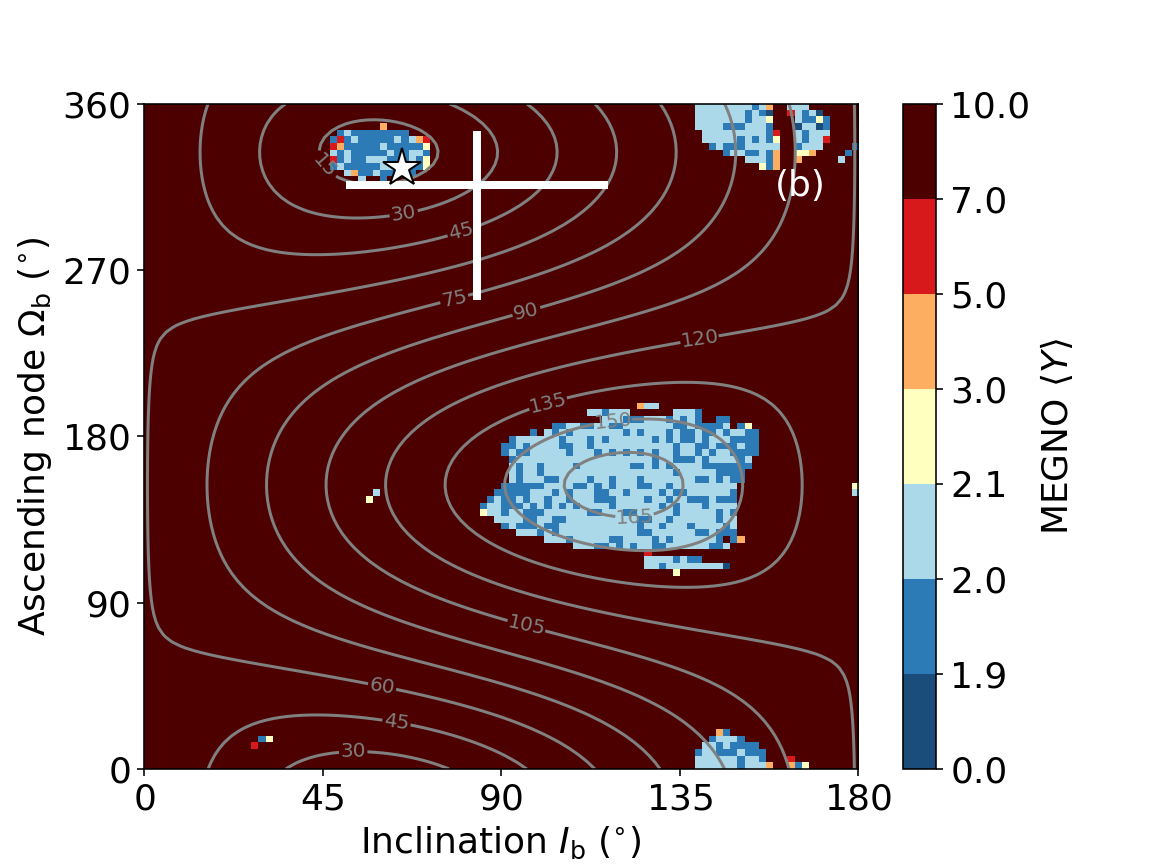}
    \caption{Mean Exponential Growth factor of Nearby Orbits (MEGNO) of the $\nu$~Oph system over 1~Myr, with the left panel (a) in the semi-major axis-eccentricity plane of companion b, the right panel (b) in the inclination-ascending node plane of companion b, with the gray curves indicating contours of constant mutual inclination. The white star with black edge represent the location of the maximum a posteriori orbital parameters, the white lines represent the posterior median values of the orbital parameters with error bars indicating the $1\sigma$ uncertainties, and colour scale represents the MEGNO value $\langle Y\rangle$ (dark blue: $0.0<\langle Y\rangle<1.9$, means fixed point or insufficient integration time; blue: $1.9<\langle Y\rangle<2.0$, lightblue: $2.0<\langle Y\rangle<2.1$, means periodic/stable; yellow: $2.1<\langle Y\rangle<3.0$, means transition; orange: $3.0<\langle Y\rangle<5.0$, red: $5.0<\langle Y\rangle<7.0$, dark red: $7.0<\langle Y\rangle<10.0$, means gradually increasing instability, with larger $\langle Y\rangle$ values indicating stronger instability.}
    \label{fig:megno}
\end{figure}
We first tested the orbital configuration by adopting the maximum a posteriori orbital parameters (i.e., $a,e,\omega,M_0,i,\Omega$). We used the \texttt{rebound} package to perform $N$-body integrations with the \texttt{IAS15} integrator. The system was integrated for 10~Myr, a timescale over which the host star is not expected to evolve significantly, with a fixed time-step of $\mathrm{d}t = 1/30$~yr. As the host star is a slow rotator and the brown dwarfs orbit at relatively large separations, stellar quadrupole and higher-order effects, as well as general relativistic corrections, can be safely neglected. We find that the mutual inclination between the two companions remains stable near its initial value ($\sim$10$^{\circ}$) throughout the integration, and that the resonant angle exhibits libration over the integration timescale, indicating a dynamically stable configuration (left panels in Fig.~\ref{fig:resonance}). This resonant and stability behaviour is similar to that found under the coplanar assumption in previous work \citep{Quirrenbach2019}. We note that assessing the stability of the system over much longer timescales is beyond the scope of this study, as stellar evolution would need to be taken into account given the relatively short lifetime of the massive host star.

Because the orbital inclinations and longitudes of the ascending nodes are not tightly constrained, we further tested the robustness of the dynamical behaviour by perturbing the initial conditions of the $N$-body integrations. Specifically, we drew the inclinations and ascending nodes from normal distributions centred on their maximum a posteriori values, with a standard deviation of $\sigma = 10^{\circ}$ and truncation at $\pm2\sigma$. In total, we performed 20 such realizations, all using the same integrator, time step, and total integration time. All simulations exhibit dynamical behaviours (librating resonant angles) similar to those obtained from the maximum a posteriori configuration. We note that this approach to perturbing the initial conditions is not rigorous but is sufficient for our purposes, given that the inclinations and ascending nodes are dynamically correlated. 

We next tested the orbital configuration by adopting the posterior median values of the orbital parameters, following the same methodology described before. This additional test was motivated by the fact that the posterior median orbital inclination of companion~b ($84^{\circ}$) differs substantially from its maximum a posteriori value ($65^{\circ}$). In this configuration, we find that the mutual inclination oscillates about its initial value of $\sim$21$^{\circ}$. Importantly, the system exhibits a near-resonant behaviour, as the resonant angle rapidly evolves into circulation (middle panels in Fig.~\ref{fig:resonance}). An additional set of 20 realizations with perturbed initial conditions yields the same circulating behaviour of the resonant angle, suggesting that the system may be dynamically chaotic on longer timescales.

We further tested the orbital configuration by adopting 20 parameter sets randomly drawn from the MCMC samples, with the true mutual inclinations of $40^{\circ}$$-$$50^{\circ}$, to check the mutual inclination mathematically drawn from MCMC. In this configuration (right panels in Fig.~\ref{fig:resonance}), the system tends to quickly evolve into chaos. In 12 out of the 20 tests, the system disrupts in a few million years, which justifies solutions with mutual inclinations exceeding 40° to be less reasonable.

To explore the long-term stability of the system over a broader parameter space, we computed the Mean Exponential Growth factor of Nearby Orbits (MEGNO, in Fig.~\ref{fig:megno}) over 1~Myr using the \texttt{rebound} package \citep{Rein2012}. We explored  the $a–e$ space and the $i–\Omega$ space of companion b, due to the weaker constraints on its $i$ and $\Omega$. For other orbital parameters, we fixed them to the maximum a posteriori solution. The locations of the maximum a posteriori and posterior median orbital parameters are denoted on the figure, showing that the maximum a posteriori orbital parameters lie in the stable area, while the inclination and ascending node of posterior median values tend to result in instability within 1~Myr. The maximum a posteriori orbital parameters are therefore more reasonable than the posterior median values, and the mutual inclination tends to be no larger than $\sim$15$^{\circ}$ according to the contour of Fig.~\ref{fig:megno}(b). This is also consistent with the unstable results obtained for the N-body simulation with initial mutual inclinations of $40^\circ-50^\circ$.

From our simulations, one general conclusion is that the system can remain in a 6:1 mean-motion resonance under conditions of moderate mutual inclination (e.g., cases of $\sim$10$^{\circ}$), whereas a larger mutual inclination preferentially leads to an unstable situation. Through observations combined with dynamics, we conclude that there is a certain mutual inclination between the two planets in this system, but it should not exceed $\sim$15$^\circ$ for stability over 1~Myr. More detailed dynamical investigations (e.g., Cao et al., in prep.) are expected to provide additional details into the system’s dynamics, but such analyses are beyond the scope of the present work.

\subsection{Brown dwarf pair systems}
\begin{figure}[t]
    \centering
    \includegraphics[width=0.98\textwidth]{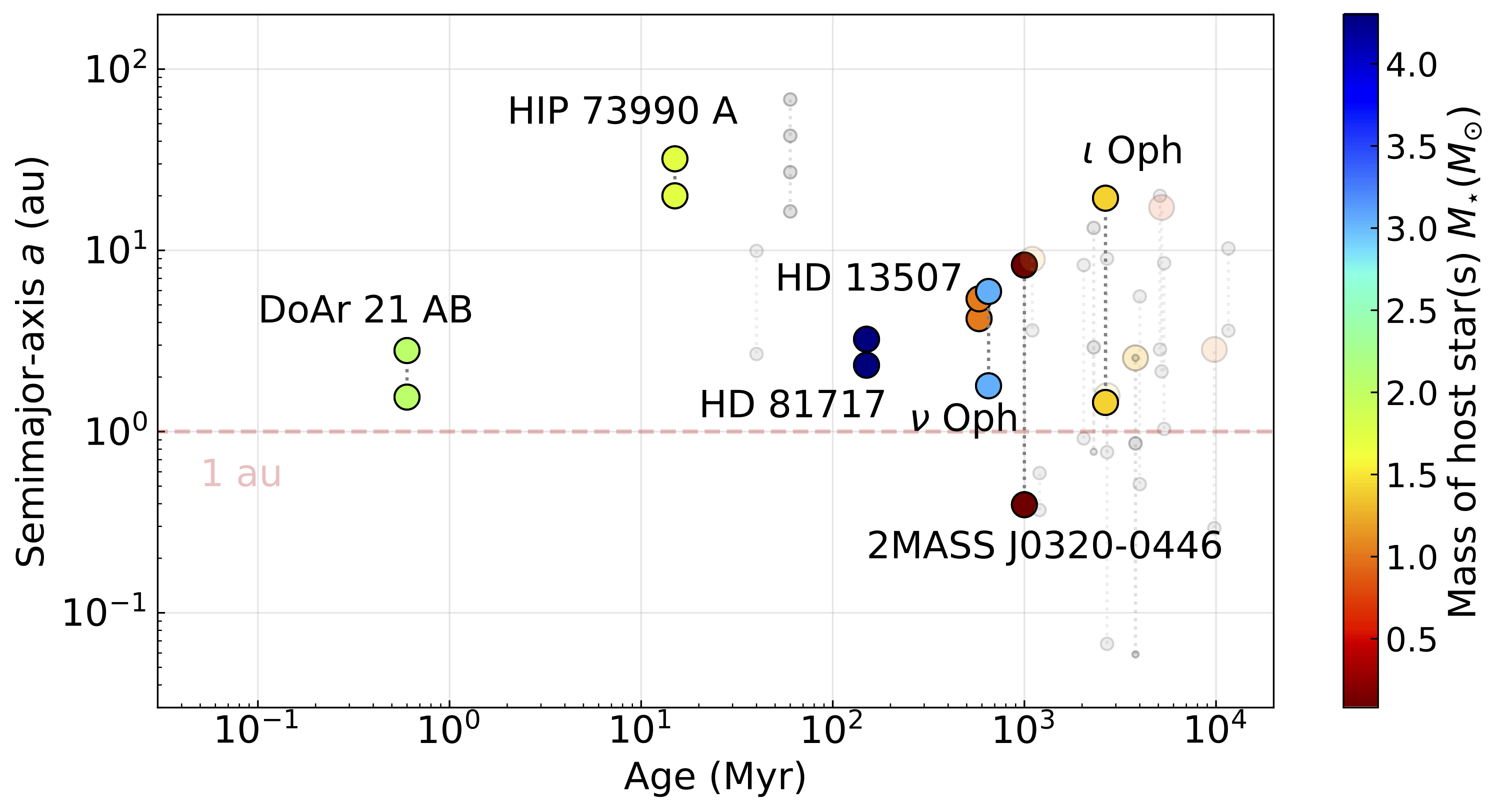}
    \caption{Orbital semi-major axis versus system age for systems hosting brown dwarf pairs. Brown dwarfs ($13\,M_{\mathrm{J}} < m < 80\,M_{\mathrm{J}}$) are shown as large, coloured circles, with the colour indicating the mass of the host star. Systems with multiple companions more massive than $5\,M_{\mathrm{J}}$ are also included and are plotted with transparent symbols; super-Jupiters ($5\,M_{\mathrm{J}} < m < 13\,M_{\mathrm{J}}$) are marked by small grey circles, while lower-mass planets are shown as grey dots. This figure shows a clear feature that brown dwarf pairs are predominantly found at orbital distances beyond 1~au from their host stars, with the sole exception of 2MASS~J0320--0446~b, which orbits a late-M dwarf at a separation of $\sim$0.3~au.}
    \label{fig:bdpair}
\end{figure}

To date, approximately 700 brown dwarf (BD; $13\,M_{\mathrm{J}} < m < 80\,M_{\mathrm{J}}$) systems have been confirmed\footnote{\url{https://exoplanet.eu}}. However, systems hosting multiple companions that include BDs are rare, and BD pairs are even more uncommon. Figure~\ref{fig:bdpair} shows the currently known BD pairs, plotted in terms of their semi-major axes versus system ages where available. Three notable characteristics emerge. First, BD pairs are predominantly found at orbital distances beyond 1~au from their host stars, with only one exception: 2MASS~J0320--0446~b, which orbits a late-M dwarf at a separation of $\sim$0.3~au. Second, BD pairs appear almost exclusively in young systems (age $<1$~Gyr), with $\iota$~Dra (age $=2.65$~Gyr) being the only known older system. Third, the host stars exhibit substantial diversity, including both single and binary systems, with stellar masses spanning a wide range from $\lesssim0.1\,M_{\odot}$ (approaching the BD mass regime) to $\gtrsim4\,M_{\odot}$.

Previous studies form statistics suggest that BDs are more likely to form in a star-like manner through rapid gravitational instability, often in relatively isolated environments (e.g., \citealt{Ma2014,Maldonado2017,Schlaufman2018}). This interpretation is physically plausible, as BDs are significantly more massive than typical gaseous planets and therefore require a larger reservoir of material from circumstellar clouds \citep{Bate2012,Luhman2012} or disks \citep{Stamatellos2009,Kratter2016}. In addition, such formation is expected to occur at larger distances from the host star, where sufficient gas is available to support the condensation of massive objects \citep{Kratter2010,Santos2017}. Given the finite amount of material in circumstellar clouds or disks, these factors---namely the high mass requirement and formation at wider separations---naturally explain both the scarcity of multiple-BD systems and the rarity of BDs forming extremely close to their host stars.

Moreover, if a BD migrates inward to a more compact configuration, their large masses can induce strong dynamical interactions with its neighbourhood \citep{Winter2022}. Such interactions can rapidly drive the system toward instability, potentially leading to the ejection of one companion. This dynamically violent process may further contribute to the low occurrence rate of BD pairs in older, more evolved systems. In addition, the diversity of host stars implies a wide range of survival environments. Circumbinary systems introduce more complex dynamical conditions in the inner regions (e.g., \citealt{Armstrong2014,Sutherland2016}), while high-mass host stars  evolve and expand on relatively short timescales (age $<1$~Gyr). Qualitatively, these effects may further inhibit the long-term survival of BDs on close orbits around their host stars.

In the future, we anticipate that forthcoming \textit{Gaia} data releases (DR4 and DR5), new direct-imaging surveys and other detection techniques, will lead to new discoveries of brown dwarf pair systems, particularly at young ages. Such findings will provide valuable observational evidence to further advance our understanding of brown dwarf formation.

\normalem
\begin{acknowledgements}
This work is supported by National Key R\&D Program of China, No. 2024YFA1611802, and the National Natural Science Foundation of China (NSFC, Grant Nos. 12588202, and 12573068). H.Y.T. appreciates the support by the EACOA/EAO Fellowship Program under the umbrella of the East Asia Core Observatories Association.
We thank Man Hoi Lee, Fei Dai, Eiichiro Kokubo, Bun'ei Sato for helpful discussion.
\end{acknowledgements}

\bibliographystyle{raa}
\bibliography{bibtex}

\newpage
\appendix 
\section{Additional figures}
Additional figures including Fig. \ref{fig:rv}, \ref{fig:mcmc_posterior_5},  \ref{fig:rv_diff} and \ref{fig:gls_residuals} are given in the appendix.
\begin{figure}[t]
\centering
\includegraphics[width=\textwidth]{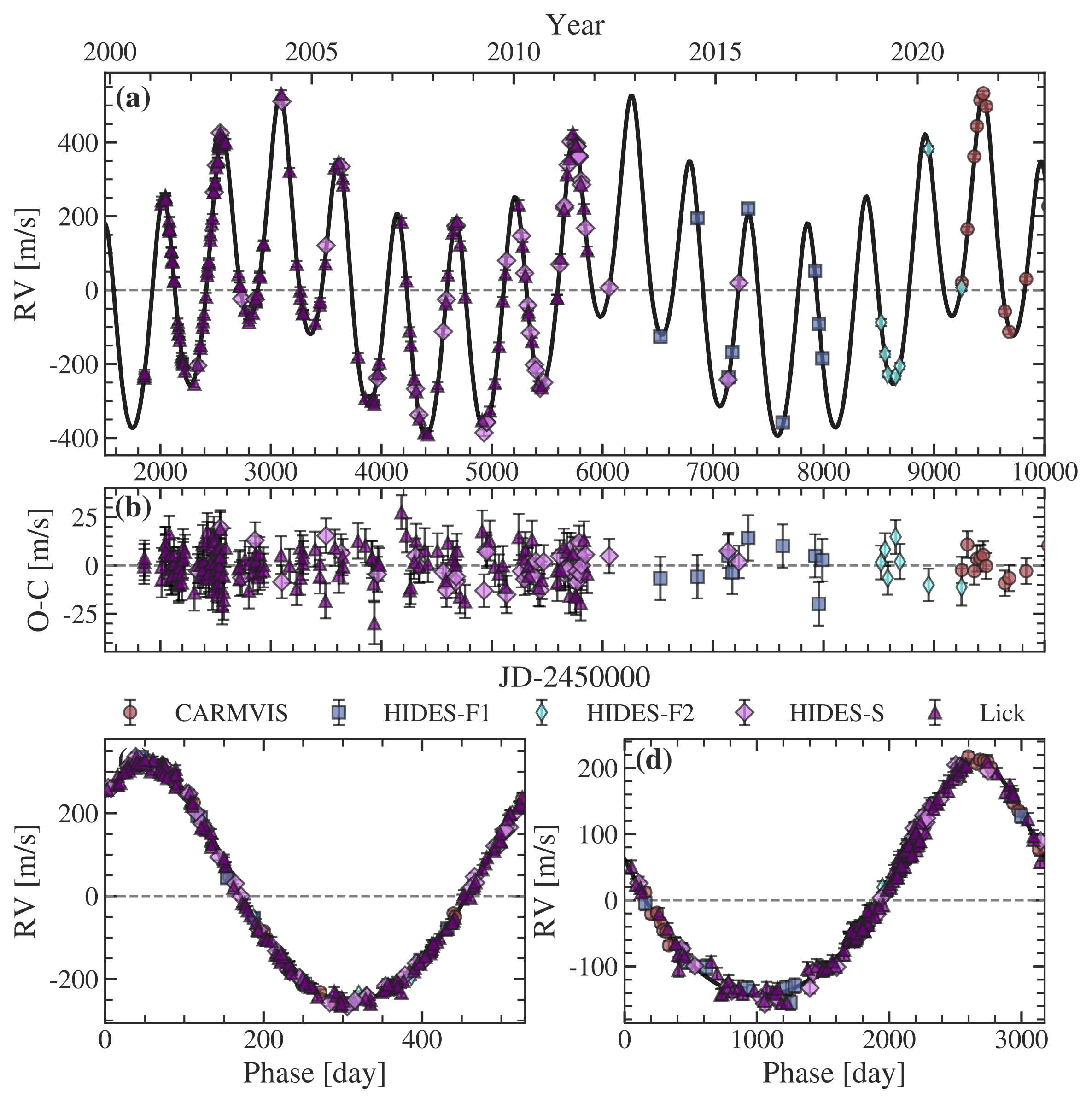}
\caption{Zoomed-in views of panels~(a) and~(b) of Fig.~\ref{fig:rvast}, shown here as panels~(a) and~(b). Panels~(c) and~(d) present the phase-folded radial-velocity signals of the two companions. As in Fig.~\ref{fig:rvast}, measurements from different instruments are indicated by different colours.}
\label{fig:rv}
\end{figure}

\begin{figure*}[t]
\includegraphics[width=\textwidth]{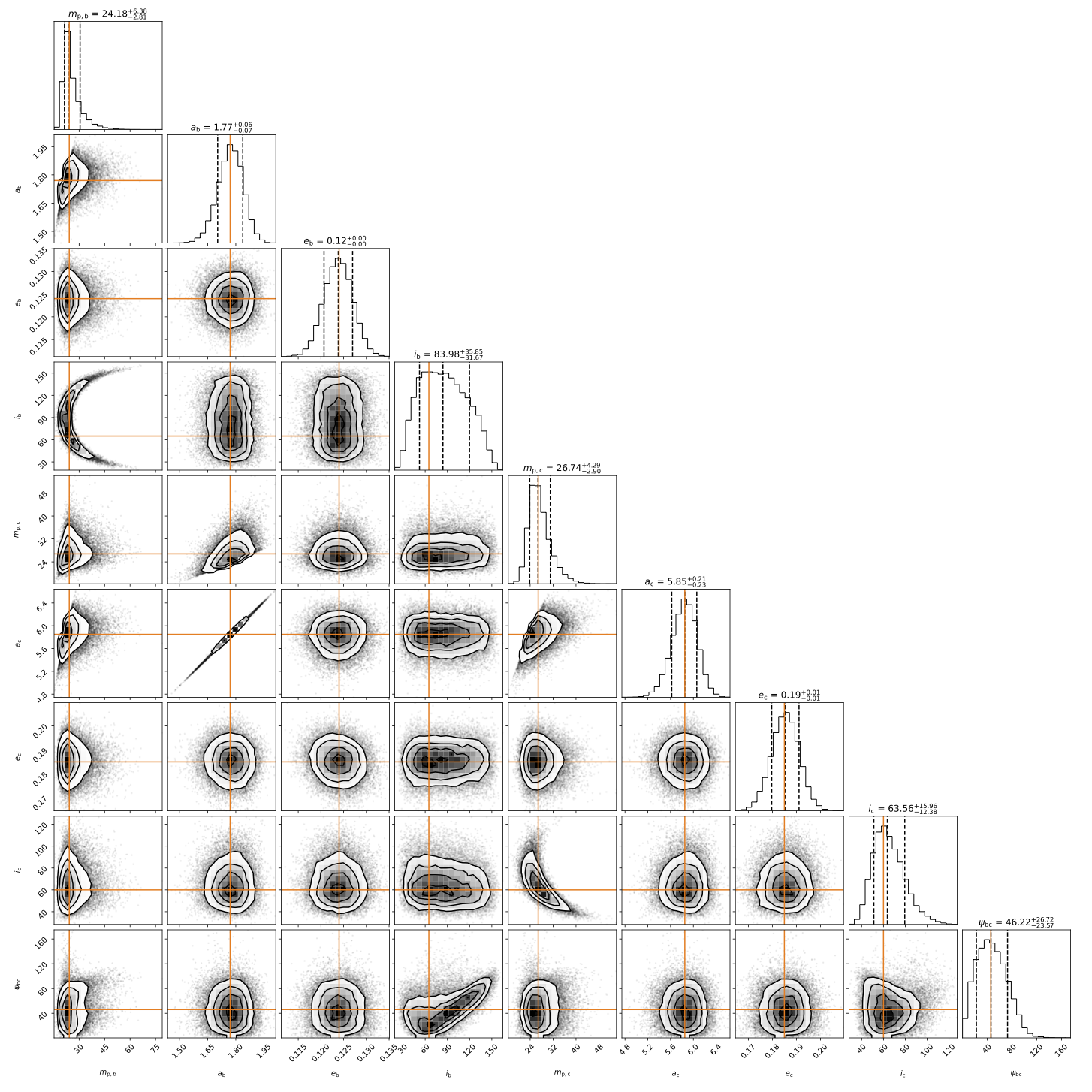}
\caption{Posterior distributions of key parameters derived from the Markov Chain Monte Carlo analysis. Shown are the companion masses $m_{\mathrm{p}}$, semi-major axes $a$, eccentricities $e$, and orbital inclinations $i$ of both companions, as well as their mutual inclination $\psi_{\mathrm{bc}}$. The median along with 16 and 84 percentiles are given in the title as well as illustrated by black dashed lines, while the values at maximum posterior are illustrated in yellow lines.}
\label{fig:mcmc_posterior_5}
\end{figure*}

\begin{figure*}[t]
\includegraphics[width=\textwidth]{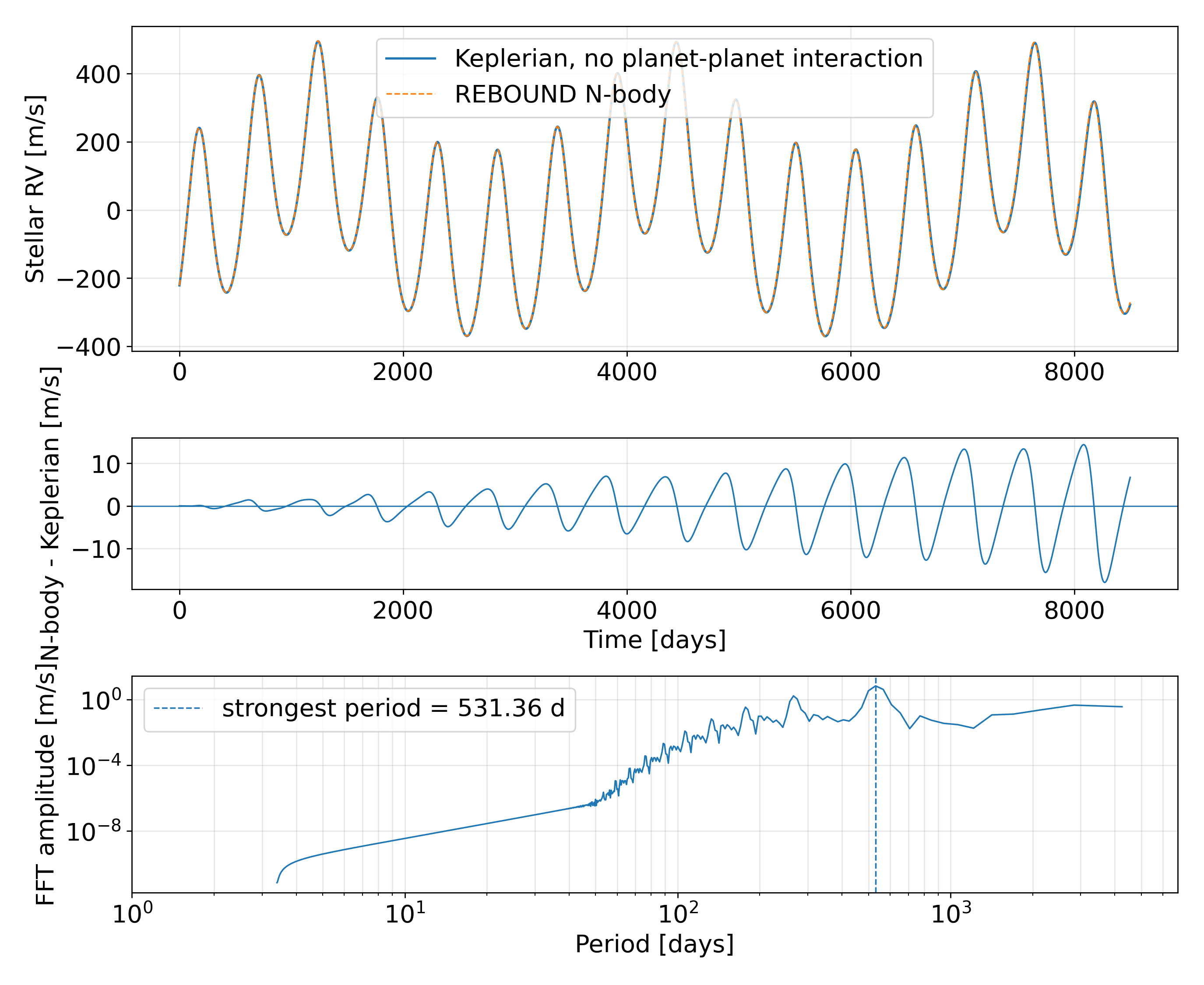}
\caption{Comparison between the Keplerian and dynamical RV models for the maximum a posteriori orbital configuration. Top: stellar RV curves predicted by the Keplerian model, in which the two companions are treated as independent Keplerian orbits, and by the full dynamical model integrated with \texttt{rebound}. Middle: RV difference between the dynamical and Keplerian models. The accumulated difference remains below $\sim$15~m~s$^{-1}$ within the current observational baseline of $\sim$8500 days. Bottom: Fourier amplitude spectrum of the RV difference. The strongest peak appears at a period of $\sim$531 days, corresponding to the dominant periodic component of the model difference over the simulated baseline.}
\label{fig:rv_diff}
\end{figure*}

\begin{figure*}[t]
\includegraphics[width=\textwidth]{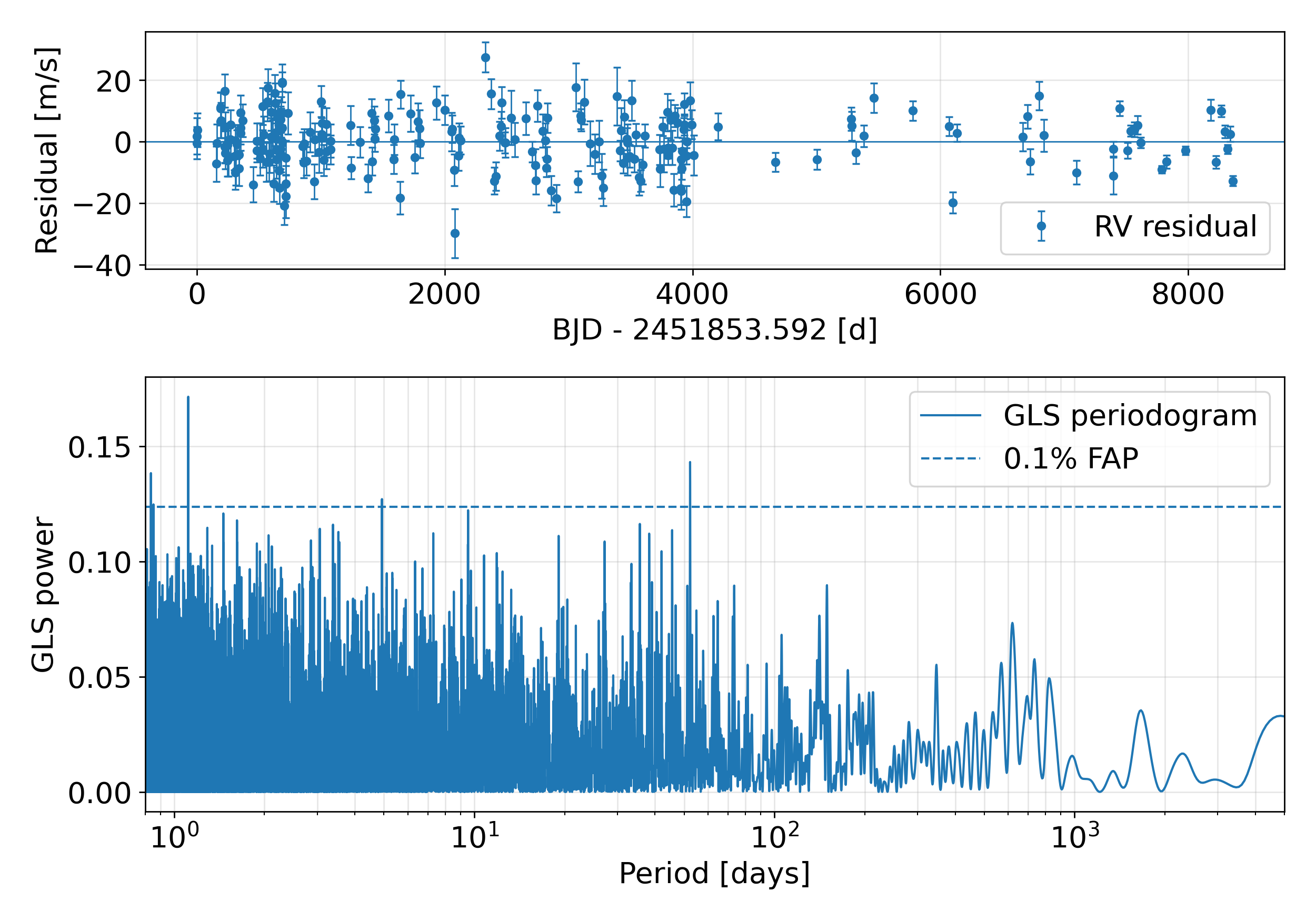}
\caption{RV residuals and their periodogram after subtracting the best-fit Keplerian model. Top: RV residuals as a function of time, where the time is shown relative to BJD 2451853.592 (first observational epoch). The residuals show no clear long-term coherent trend over the observational baseline. Bottom: Generalised Lomb--Scargle periodogram of the RV residuals. The dashed horizontal line denotes the 0.1\% false-alarm probability level. No significant additional long-period signal is detected in the residuals, while the remaining short- to intermediate-period peaks are likely associated with unresolved stellar variability, instrumental noise, or sampling aliases.}
\label{fig:gls_residuals}
\end{figure*}

\end{document}